\documentclass[twocolumn,prl,aps,groupedaddress,floats,showpacs,final,superscriptaddress]{revtex4-1}
\usepackage[latin3]{inputenc}
\usepackage[makeroom]{cancel}
\usepackage{graphicx}
\usepackage{amsmath}
\usepackage{amsfonts}
\usepackage{amssymb}
\usepackage{chngcntr}
\usepackage{color}
\usepackage[left=2cm,right=2cm,top=2cm,bottom=2cm]{geometry}

\usepackage{graphicx}
\usepackage{dcolumn}
\usepackage{bm}
\usepackage{simplewick}
\usepackage{array}
\usepackage{appendix}

\RequirePackage[
   hyperindex,colorlinks,bookmarksnumbered,
   plainpages=true,pdfstartview=FitH]{hyperref}
\hypersetup{linkcolor=blue,urlcolor=blue,citecolor=blue}
\usepackage{hyperref}

\definecolor{purple}{rgb}{0.5,0,0.6}

\usepackage{ulem}

\begin{document}

\date{\today}
\title{Andreev probing of Cooper-pair flying qubit}

\author{S. Park}
\affiliation{Center for Theoretical Physics of Complex Systems,
Institute for Basic Science (IBS), Daejeon 34126,
Republic of Korea}

\author{L. Y. Gorelik}
\affiliation{Department of Physics, Chalmers University of
Technology, SE-412 96 G{\" o}teborg, Sweden}

\author{S.I.Kulinich}
\affiliation{B. Verkin Institute for Low Temperature Physics and
Engineering of the National Academy of Sciences of Ukraine, 47
Prospekt Nauky, Kharkiv 61103, Ukraine}

\author{H. C. Park}
\affiliation{Department of Physics, Pukyong National University, 
Busan 48513, Republic of Korea}

\author{C.Kim}
\affiliation{Quantum Information Center, Korea Institute of Science and Technology, 
Seoul 02792, Republic of Korea}

\author{R. I. Shekhter}
\affiliation{Department of Physics, University of Gothenburg, SE-412
96 G{\" o}teborg, Sweden}
\date{\today}
\begin{abstract}
We propose a nanomechanical device which can actuate 
and probe a flying qubit that can be used to facilitate quantum information transfer 
over a long distance. The flying qubit is formed by a movable Cooper pair box (CPB) 
consisting of a superconducting dot and a bulk superconductor which are entangled 
by removing the Coulomb blockade of Cooper pair tunneling electrostatically. 
We suggest that flying qubit states formed on movable CPB can be observed in electron 
transport to a normal electrode via Andreev reflections. The charge transfer due to 
periodic mechanical motion of CPB leads to nonzero current at zero voltage 
and its coherence can be identified through oscillatory dependence of the current on a gate voltage.
\end{abstract} \maketitle
{\it Introduction.---}
Micro- and nano-fabrication techologies have evolved up to the point where we can reproduce the circuit design with nanoscale precision and control its fabrication yield in a reasonable level \cite{chulki}. Nanoelectromechanical systems (NEMS) have been developed for the last two decades along this stream of technological achievements. 
At the heart lies nanomechanical degrees of freedom coupled to electron transport in 
the quantum mechanical regime \cite{cleland,ekinci,mesonm}.       
The sequential electron tunneling by the mechanical oscillation of a metallic grain, 
coined as ``electron shuttle'', was proposed in 1998 \cite{shuttle}, offering a noble 
platform for spatial transportation of electrons. 
It was naturally followed by its generalization towards 
coherent transfer of Cooper pairs with a movable superconducting grain \cite{CPB shuttle}.    
Nanomechanically-assisted transports of electric charge \cite{chulki2012}, electronic spin \cite{fedorets2005} and heat \cite{vikstrom2016,morell2019} have been suggested as well. 
Composite mesoscopic structures based on nanomechanical resonators have recently come into 
the focus of modern research due to the possible acoustic implementation of quantum 
communication systems \cite{sawflying,injection,setqd}. 

Non-local information, such as superposition and entanglement, that encoded on 
electronic states located at different movable parts of the NEMS has brought a new 
paradigm in NEM-transduced electronics, so called a nanomechanical processor for quantum information. Transfer or manipulation of quantum information can be achieved as suggested for electric charge, spin, and heat in standard NEMS. 
Namely, the concept of ``flying qubits'' can be mechanically implemented with intriguing perspectives in mechanically active electric weak links \cite{shuttle,Shekhter2003}. 
Flying qubits have been realized by using single photons within superconducting circuits \cite{Houck2007} or propagating electrons along a one-dimensional channel \cite{Yamamoto2012}. 
To open a new avenue of the nanomechanical flying qubit and address limitations of the 
aforementioned platforms, a theoretical evidence for such an implementation should be proposed in the first place. 

In this work, we suggest a nanomechanical device in which a single Cooper pair formed by 
a superposition between charge states can be transported, actuated, and probed. Since the 
charge states formed in a movable Cooper pair box (CPB) constituting a qubit, we call it a  
{\it Cooper-pair flying qubit}. We adopt Andreev reflection, constraining the tunneling of a 
normal metal electrons into moving CPB, to access the qubit states. We found that a 
nonzero current can be pumped in such a system even at zero driving voltage. 
The magnitude and direction of such current is determined by nanomechanical motion of CPB and 
states of a movable qubit. The later discloses its oscillatory dependence on electrostatic gate potential. It turned out that such oscillating quantum states at high frequency manifests 
under its inverse proportionality to mechanical frequency of the CPB. 

We highlight the main points of our work. 
Our main result Eq. \eqref{CurrentFF} shows the strong spatial non-locality of the current response to the electrostatic gate potential, 
manifesting the quantum ``rigidity'' during its mechanical motion and the 
preservation of quantum coherence in the flying qubit. We also note that the normal electrode 
as a tool to probe the current can be regarded as a dissipative environment and causes the dissipation of the charge states. Thus, our work addresses the question how the dissipation 
affects the flying qubit.

\begin{figure}
\includegraphics[width=1.0\columnwidth]{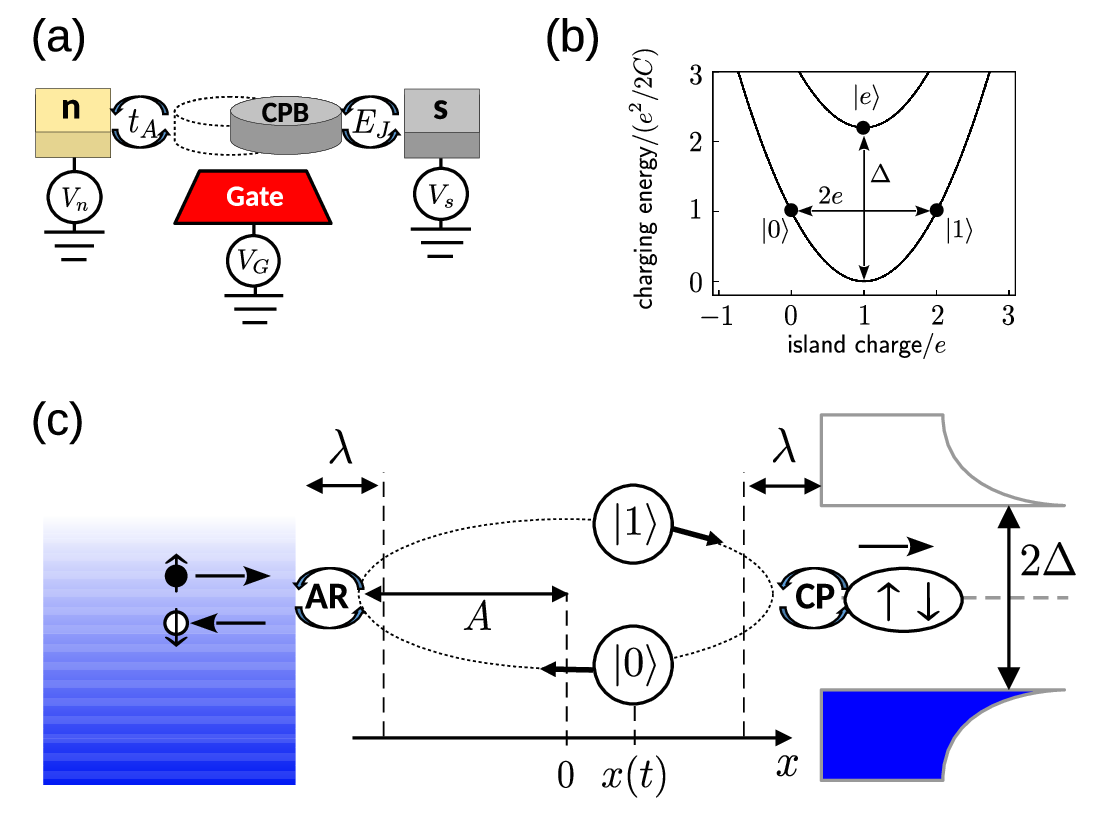}
\caption{A schematic of the nanomechanical device for the Cooper-pair flying qubit. 
(a) The device consists of a superconducting island denoted by CPB which moves periodically between n- and s-electrodes. Voltages are applied to the electrodes ($V_s$ and $V_n$) and to the gate at $x=0$ ($V_G$) in the middle.
(b) Charging energy as a function of the charge of the CPB. 
The energy branches of an even and odd number of electrons are differ by the superconducting gap $\Delta$. The gate voltage $V_G$ can be adjusted to bring the charge states ($|0\rangle$ and $|1\rangle$) in the even branch to degenerate and ensure that the Coulomb blockade is removed. 
(c) The CPB, whose position is denoted by $x(t)=A \sin \omega t$ with a frequency of $\omega$, is tunnel coupled with the s-electrode through a Josephson coupling $E_J$ at $x= A$ accompanying Cooper pair tunneling (CP) and the n-electrode via Andreev reflection (AR) of amplitude $t_A$ at $x=-A$. The position dependence of the tunneling is characterized by the tunneling length $\lambda$}\label{fig1}

\end{figure}

{\it The system.---}
A sketch of the system under consideration is shown in Fig. \ref{fig1}.
It consists of superconducting (s) and normal (n) electrodes
separated by a distance much greater than the tunneling length
$\lambda$. A tiny superconducting island, whose coordinate is
determined by $x(t)$, performs harmonic oscillations with amplitude
$A$ and frequency $\omega$,($x (t)= A\sin\omega t$) periodically
approaching electrodes at a distance of the order of $\lambda$.
Thus, the superconducting island can only have a tunnel connection
with one of the electrodes at a given time. Gate electrode located
near the central value $x=0$ controls the electrostatic potential on
the island as it passes by. We assume that the superconducting
energy gap (on the island and s-electrode) $\Delta$ is much greater
than the charging energy of the island $E_{C}(x)=e ^{ 2 } /2C(x)$ ($
C(x) $ is the island's capacitance), and the energy of the Josephson
connection between the island and a superconductor $E_{J}(x)$.
This allows us to consider the subsystem consisting of the island and
s-electrode as a two-level system - Cooper pair box, whose states
are represented by a superposition of two states: $|0\rangle=(0,1)^
{T }$ - the ground state of the neutral island, and
$|1\rangle=(1,0)^{T}$ - the ground state of the island with two
extra electrons \cite{matveev1993,Nakamura,Bludh}.

To describe the dynamics of such a system, we use the following
Hamiltonian:
\begin{eqnarray}\label{hamiltonian}
H= H_{0}+H_{A},\,\,\,\, H_{0}=H_{CPB}+H_{n}.
\end{eqnarray}
Here the Hamiltonian $H_{CPB}$ describes the CPB
\begin{eqnarray}\label{hamiltonian}
H_{CPB}= E_{J}(x(t))\hat{\sigma}_{1}+ E_{Q}(x(t))\hat{\sigma}_{3},
\end{eqnarray}
where $\hat{\sigma}_{i=1,2,3}$ are Pauli matrices acting on the two-level system, 
$E_{J}(x)=E_{J}\exp\{-2(A-x)/\lambda\}$ is the energy of the
position depended Josephson coupling between island and s-electrode, where 
$\lambda$ is the tunneling length and $E_{J}\approx \Delta\left(2e^{2}/h\right) R_{s}\sim
10^{-5}\div10^{-6}eV$ ($R_{s}$ is a minimal resistance of the tunnel
junction between the island and s-electrode when both are in a normal
state). $E_{Q}(x)=E_{C}(x)-eV_{s}+e\alpha(x)V_{G}$ is a charging
energy where $V_s$ is a voltage applied to the s-electrode and
$\alpha(x)V_{G}$ is an electrostatic potential induced by the gate
electrode (see Fig. \ref{fig1}) near the central point $x=0$.

The Hamiltonian $H_n$ describes the n-electrode,
\begin{eqnarray}\label{03}
&& H_n=\sum_{k\kappa}(\varepsilon_k^{n}-\mu) a_{k,\kappa}^\dag a_{k,\kappa},
\end{eqnarray}
Here $a^\dag_{k,\kappa} (a_{k,\kappa})$ are creation (annihilation)
operators of the electron with quantum number $k$ and spin
projection $\kappa$ in the n-electrode, $\varepsilon_k^{n}$ are
the single electron energies of the n-electrode.

And the last term $H_{A}$ describes the Andreev-type electrons
exchange between the n-electrode and superconducting island, when
island is nearby on the distance of the tunneling length,
\begin{equation}\label{An}
    H_{A}=\mathrm{t}_{\mathrm{A}}(x(t))\sum_{k,k'}\left(a^\dag_{k,\kappa}a^\dag_{k',-\kappa}\hat{\sigma}^{-}+a_{k,\kappa}a_{k',-\kappa}\hat{\sigma}^{+}\right).
\end{equation}
Here $\hat{\sigma}^{\pm}=(\hat{\sigma}_{1}\pm i\hat{\sigma}_{2})/2$,
$\mathrm{t_{A}}(x)=\mathrm{t_{A}}\exp\{-2(A+x)/\lambda\}$,
$\mathrm{t_{A}}\approx D(R_{0}/R_{n})$, $D$ is a band
width,$R_{0}=(h/e^{2}$ is an inverse  quantum conductance, and
$R_{n}$ is a minimal resistance of the tunnel junction between the
n-electrode and island when the latter is in a normal state.

The state of the CPB is described by $2\times 2$ matrix
$\hat{\varrho}$ with elements $\varrho_{i,i'}=Tr_{n}\langle
i|\hat{\varrho}_{tot}|i'\rangle-\frac{1}{2}\delta_{i,i'}$ where
$i=0,1$, $\hat{\varrho}_{tot}$ is the total density matrix and
$Tr_{n}$ denotes a trace over  the normal metal's states. Using
reduced density matrix approach (see SM~\cite{Suppl}) we get
the following equation describing time evolution of $\hat{\varrho}$,
\begin{align}
\frac{d\hat{\varrho}(t)}{dt}&=-i\frac{1}{\hbar}[H_{CPB}(t),\hat{\varrho}(t)]-\check{\hat{\mathfrak{L}}}(\hat{\varrho}),\label{mastereq}
\\
\check{\hat{\mathfrak{L}}}(\hat{\varrho})&=\gamma(t)\left(\hat{\varrho}(t)
+\frac{1}{2}\hat{\sigma}_{3}\{\hat{\sigma}_{3},\hat{\varrho}\}
-\hat{\sigma}_{3} f_{eq}\right). \label{Lindbladian}
\end{align}
Here $f_{eq}=-\tanh(E_Q^{(n)}/k_{B}T)$,
$\gamma(t)=\Gamma\exp\{-4A(1+\sin\omega t)/\lambda\}$, $\Gamma=
2\pi\hbar^{-1} t^{2}_{A}\nu^{2}E_Q^{(n)}\coth(E_{Q}^{(n)}/k_{B}T)$,
 $E_{Q}^{(n)}=E_{Q}(-A)-e\Delta V$, $\Delta V$ is a voltage
difference between normal and superconducting electrodes, and $\nu$
is electronic density of states. Getting the Lindbladian term
$\check{\hat{\mathfrak{L}}}(\hat{\rho})$ on the right side of Eq.~\eqref{Lindbladian} we neglect the small renormalization ($\sim
t^{2}_{A}\nu^{2}<<1$) of the charging energy $E_{Q}^{(n)}$, induced
by the interaction with n-electrode (see SM~\cite{Suppl}).

\begin{figure}
\includegraphics[width=0.9\columnwidth]{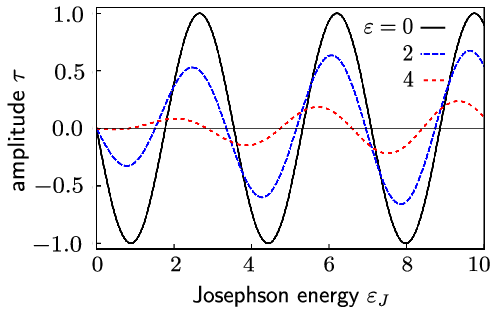}
\caption{The amplitude of the Cooper pair tunneling $\tau$, occurring at $x=A$, as a 
function of the normalized Josephson energy $\varepsilon_J$. It is computed numerically by solving Eq.~\eqref{Us}. The plot shows that as the charging energy $\varepsilon$ increases, the maximum of the tunneling amplitude decreases.}\label{fig2}
\end{figure}

{\it Stationary state and average current.---}
Our analysis shows that, regardless of the initial state
$\hat{\hat{\varrho}}(0)$, the time evolution of the density matrix enters
stationary periodic mode
$\hat{\varrho}_{st}(t)=\hat{\varrho}_{st}(t+T_{0})$ with the time period 
of the mechanical motion $T_{0}=2
\pi/\omega$ (see SM~\cite{Suppl}). To analyze this regime, it
is convenient to present time interval $(nT_{0},(n+1)T_{0})$ on a
sum of two intervals $(nT_{0},(n+1/2)T_{0})$ and
$((n+1/2)T_{0},(n+1)T_{0})$. Considering that during the first
interval $\gamma(t)=0$, and hence the time evolution of the CPB
purely unitary, we get the following relation between
$\hat{\varrho}_{1}\equiv\hat{\varrho}_{st}(nT_{0})$ and
$\hat{\varrho}_{2}\equiv\hat{\varrho}_{st}((n+1/2)T_{0})$.
\begin{align}
  \hat{\varrho}_{2}&= e^{-i\Phi_{1}\hat{\sigma}_{3}}\hat{U}_{s}\hat{\varrho}_{1}\hat{U}_{s}^{\dag}e^{i\Phi_{1}\hat{\sigma}_{3}}, \label{FirstIn}\\
  \hat{U}_{s} &=
  \mathcal{T}_{t_{1}}\exp\{-i\varepsilon_{J}\int^{\infty}_{-\infty}dt_{1}
 e^{-t^{2}_{1}}(\hat{\sigma}_{1}\cos\varepsilon t_{1}
 +\hat{\sigma}_{2}\sin\varepsilon t_{1})\} \nonumber\\
   &=e^{-i\Phi_{s}\hat{\sigma}_3/2}(\rho+i\tau\hat{\sigma}_{1})e^{-i\Phi_{s}\hat{\sigma}_{3}/2}. \label{Us}
\end{align}
Here $\Phi_{1}=\hbar^{-1}\int^{T_{0}/2}_{0}dt E_{Q}(A\sin\omega t)$,
$\varepsilon_{J}=\pi^{1/2}\sqrt{\frac{\lambda}{A}}\frac{E_{J}}{\hbar\omega}$,
$\varepsilon=\pi^{1/2}\sqrt{\frac{\lambda}{A}}\frac{E_{Q}(A)}{\hbar\omega}$, $\mathcal{T}_{t_1}$ is the time-ordering operator. The
parameter $\tau^{2}$ has a simple physical meaning; it gives the
probability of changing number of Cooper pairs in the island from the neutral state $|0\rangle$ to the charge state $|1\rangle$, or vice versa, when the island approaches and then moves away from the s-electrode.
The amplitude $\tau$ as function of $\varepsilon_{J}$ in general
case at different $\varepsilon$ is presented in Fig.~\ref{fig2}.

Average charge transferred to a superconductor $Q$ on the one period
is determined by the equation,
\begin{equation}\label{charge}
    Q=\int^{\frac{T_{0}}{2}}_{0}dt \mathrm{Tr}\hat{j}(t)\hat{\rho}(t)=e\mathrm{Tr}\hat{\sigma}_{3}(\hat{\rho}_{2}-\hat{\rho}_{1})
\end{equation}
where $\hat{j}(t)=(2eE_{J}/\hbar)\hat{\sigma}_{2}$ is a current
operator. 

We found the solution of Eq.~\eqref{mastereq} formulated above with the periodic boundary condition, and found the following equation for $Q$, see SM \cite{Suppl},
\begin{equation}
 Q=-e\,\text{tanh}\left(\frac{E^{(n)}_Q}{k_B T}\right)\frac{\tau^2 \sinh\bar{\Gamma}}{\cosh\bar{\Gamma}-\rho^{2}\cos2\Phi}. \label{ChargeQ}  
\end{equation}
Here $\Phi=\Phi_0+2\pi V_G/V_0$ where $\Phi_0 = \hbar^{-1} \oint
dtE_{Q}(x(t))+\Phi_{s}$ and $V_{0}\approx (\pi\hbar\omega/e)(A/d)$ with 
a characteristic length of the gate electrode $d$. The parameter $\bar{\Gamma}=\pi^{1/2}\sqrt{\gamma/2A}\Gamma/\omega\ll 1$ which gives the probability of changing 
number of Cooper pairs in the island from $|0\rangle$ to $|1\rangle$, or vice versa, when the island approaches and then moves away from the n-electrode. In such a limit the average current
$I=Q/T_{0}$:
\begin{equation}\label{CurrentFF}
    I=I_{A}\frac{\tau^{2}}{\tau^{2} +2\rho^{2}\sin^{2}\Phi} =I_{A}\tau^{2}\sum_{n=0}^{\infty}\left(\rho^{2}\cos(2\Phi)\right)^{n}
\end{equation}
Here
$I_{A}=-2e\sqrt{\frac{\pi\lambda}{A}}\left(\frac{G_{n}}{G_{0}}\right)^{2}\left(\frac{E_{Q}^{n}}{\hbar}\right)$. 
In the  opposite limit
$\bar{\Gamma}\gg 1$ the average charge transferred to the
superconductor $Q=ef_{eq}\tau^{2}$ does not depend on quantum phase
$\Phi$ - pure stochastic regime.

\begin{figure}
\includegraphics[width=1.0\columnwidth]{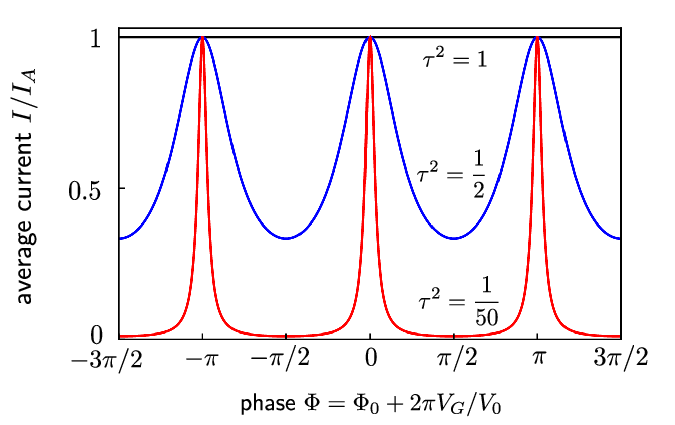}
\caption{Average current $I$ in a stationary regime as a function of a gate voltage $V_G$.  The resonant peaks correspond to the constructive interference of coherent pumping of Cooper pairs between the Cooper-pair flying qubit and the electrodes. As the Cooper pair tunneling probability $\tau^2$ decreases, the width of the resonant peak decreases. }\label{fig3}
\end{figure}

{\it Discussion.---}
The average current is proportional to $E^{(n)}_Q$ the difference 
between Coulomb energy $e^{2}/C(A)$ and the shift of the chemical potential 
arising from the voltage difference $V_{n}-V_{s}$ applied to the n- and s- 
electrodes. And it varies periodically on the gate voltage $V_G$ through the phase $\Phi$.
The normalized current $I/I_{A}$ as a function of $\Phi$ at
different values of $\tau$ is presented in Fig.~\ref{fig3}, 
exhibiting sharp resonant peaks at $V_G = V_0 (n/2 - \Phi_0/2\pi)$ with the integer $n$.
The resolution of the gate voltage required to observe the resonance peak is 
$\Delta V_G \ll V_0/2 \approx 1\mu$V with $\omega/2\pi = 10$ GHz, which is feasible in current experimental techonology.  
The temperature should be smaller than the separation between the resonance peaks 
$e V_0/2 = (\pi A/2 d)\hbar \omega > k_B T$.

The NEM-based Cooper-pair flying qubit remains isolated from an environment 
during the transportation, ensuring a long coherence time $T_2$. Moreover, in comparison 
with flying electrons implemented in two-dimensional semiconductors with $T_2 \approx$ 1-100 ns, the superconducting qubit has in general longer $T_2 \approx$ 1-100 $\mu$s \cite{Heinrich2021}. 

Eqs.~\eqref{ChargeQ} and \eqref{CurrentFF} are the main results of our analysis. 
It presents the electrical current through
the device as a result of nanomechanical pumping of Cooper pairs. As
such, the finite current exists even in the absence of the bias voltage.
Two couplings occurring at a distance from each other are responsible 
for such a pumping. One is tunneling coupling of the CPB with normal
metal reservoir which induces noncoherent, stochastic tunneling events
affecting the population of the two level system. Such perturbation
serves as a source of dissipation constraining a life time of the
flying qubit. It is a typical element for the classical pumping
device. Another perturbation shifted in time as compared with the
first one is scattering of two level qubit state caused by
tunneling coupling with superconductor. Such scattering involves
quantum state on the CPB and BCS state in the the quantum evolution
and represents the perturbation typical for quantum pumping. As a
result current flow through our device should be viewed as an
interplay of classical and quantum pumping elements. As a
consequence, dissipationfull phenomenon is still dependent on quantum
mechanical phase resulting in mechanically assisted Andreev
current. 

{\it Conclusion.---}
We found that the Andreev reflection can probe a Cooper-pair flying qubit with 
the finite current even in the absence of the bias voltage. The direction of 
the current depends on the difference between the Coulomb energy and the chemical 
potential of the electrodes ($e^2/C-\mu$); if positive, Cooper pairs transfer 
from the superconducting to the normal electrode, if negative, vice versa. 
The magnitude of the current shows periodic dependence on the gate voltage. 
At small values of the parameters $\bar{\Gamma} \ll 1$ and $\tau^2 \ll 1$, which control 
exchange of Cooper pairs with normal and superconducting electrodes, respectively, the current 
has sharp resonant peaks (see Fig. \ref{fig3}). The latter is 
fairly reminiscent of the Fabry-Perot resonance. If the parameter $\bar{\Gamma} \gg 1$ is 
proportional to $\tau^2$ and does not depend on $\bar{\Gamma}$ and $V_G$.      

S.P. acknowledges the financial support from the Institute for Basic Science (IBS) in the Republic of Korea through the project IBS-R024-Y4. This work was partly supported by the KIST institutional program (2E32801) funded by the Korea Institute of Science and Technology and Institute for Information \& communications Technology Planning\&Evaluation (IITP) grant funded by the Korea government(MSIT) (No.RS-2023-00230717).


\clearpage

\setcounter{equation}{0}
\setcounter{figure}{0}
\renewcommand{\theequation}{S\arabic{equation}}
\renewcommand{\thefigure}{S\arabic{figure}}
\renewcommand*{\citenumfont}[1]{S#1}
\renewcommand*{\bibnumfmt}[1]{S#1}

\widetext
\begin{center}
\textbf{\large Supplemental material for ``Andreev probing of Cooper-pair flying qubit''}

\bigskip

S. Park$^1$, L. Y. Gorelik$^2$, S. I. Kulinich$^3$, H. C. Park$^4$, C. Kim$^5$, and R. I. Shekhter$^6$
\break \\
$^{\it{1}}$\textit{Center for Theoretical Physics of Complex Systems,\\
Institute for Basic Science (IBS), Daejeon 34126,
Republic of Korea}\\
$^{\it{2}}$\textit{Department of Physics, Chalmers University of
Technology, SE-412 96 G{\" o}teborg, Sweden}\\
$^{\it{3}}$\textit{B. Verkin Institute for Low Temperature Physics and
Engineering of the National \\ Academy of Sciences of Ukraine, 47
Prospekt Nauky, Kharkiv 61103, Ukraine}\\
$^{\it{4}}$\textit{Department of Physics, Pukyong National University, 
Busan 48513, Republic of Korea}\\
$^{\it{5}}$\textit{Quantum Information Center, Korea Institute of Science and Technology, 
Seoul 02792, Republic of Korea}\\
$^{\it{6}}$\textit{Department of Physics, University of Gothenburg, SE-412
96 G{\" o}teborg, Sweden}
\end{center}

\section*{A. Reduced density matrix approach}
We derive Eq.~\eqref{mastereq} in the main text by solving the equation, 
\begin{eqnarray}\label{Liouville}
i\hbar \partial_t \hat{\varrho}_{tot}(t) &=& \left[ H_0, \hat{\varrho}_{tot}(t) \right]
+\left[ H_A(t), \hat{\varrho}_{tot} (t) \right],
\end{eqnarray} 
where $\hat{\varrho}_{tot}$ is the total density matrix. Let us transform $\hat{\varrho}_{tot}$ 
to the interaction picture $\hat{\varrho}_{I}$ as 
$\hat{\varrho}_{tot}(t) = e^{-i H_0 t/\hbar} \hat{\varrho}_{I}(t) e^{i H_0 t/\hbar}$.
Substituting it into Eq.~\eqref{Liouville}, we obtain 
\begin{eqnarray}\label{Interaction}
\partial_t \hat{\varrho}_{I}(t) = -\frac{i}{\hbar} 
e^{i H_0 t/\hbar} \left[ H_{A}(t), \hat{\varrho}_{tot}(t)\right] e^{-i H_0 t/\hbar},
\end{eqnarray}  
leading to 
\begin{eqnarray}\label{rho_i}
\hat{\varrho}_{I}(t) = \hat{\varrho}_{I}(-\infty) 
-\frac{i}{\hbar} \int^{t}_{-\infty} dt' \,
e^{i H_0 t'/\hbar} \left[ H_{A}(t'), \hat{\varrho}_{tot}(t')\right] e^{-i H_0 t'/\hbar} e^{-\delta (t-t')/\hbar},
\end{eqnarray}  
where the small positive number $\delta$ is introduced to ensure proper convergence. We substitute 
this form of $\hat{\varrho}_{I}$ into the commutator with $H_A$ in Eq.~\eqref{Liouville}, 
\begin{eqnarray}\label{Liouville2}
\partial_t \hat{\varrho}_{tot}(t) &=& -\frac{i}{\hbar}\left[ H_0, \hat{\varrho}_{tot} (t) \right] 
-\frac{1}{\hbar^2}\left[ H_A(t), \int^{t}_{-\infty} dt' \,
e^{i H_0 t'/\hbar} \left[ H_{A}(t'), \hat{\varrho}_{tot}(t')\right] e^{-i H_0 t'/\hbar}
 e^{-\delta (t-t')/\hbar} \right]. 
\end{eqnarray} 
Here we omitted the term associated with $\hat{\varrho}_{I}(-\infty)$ due to the fact that 
$Tr_n \left[ H_A, \hat{\varrho}_I(-\infty)  \right] =0$ and thus does not contribute 
to Eq.~\eqref{mastereq}. We approimate $H_A(t')$ in the integrand to $H_A(t)$ as 
the Andreev-type electron exchange between the normal electrode and the superconducting island 
is assumed to be maximum at $t$. Also, we write the density matrix as a product state 
$\hat{\varrho}_{tot}(t) = \hat{\varrho}(t) \otimes \hat{\rho}_n$ where $\hat{\varrho}(t)$ and $\hat{\rho}_n$ are 
density matrices for the CPB and the normal metal, respectively. $\hat{\varrho}(t')$ may be expressed as 
\begin{eqnarray}\label{rho-cpb} 
\hat{\varrho}(t') \approx e^{i H_{CPB}(t-t')/\hbar} \hat{\varrho}(t) e^{-i H_{CPB}(t-t')/\hbar}
\end{eqnarray}    
to keep the time evolution of the phase factor associated with the charging energy. Then, 
the second term on the right hand side in Eq.~\eqref{Liouville2} can be written as 
\begin{eqnarray}\label{Liouville3}
&&-\frac{1}{\hbar^2}\left[ H_A(t), \int^{t}_{-\infty} dt' \,
e^{i H_0 t'/\hbar} \left[ H_{A}(t'), \hat{\varrho}_{tot}(t')\right] 
e^{-i H_0 t'/\hbar} e^{-\delta (t-t')/\hbar} \right] \\
&&=-\frac{t^2_{\text{A}}}{\hbar^2}\sum_{k_i} 
\Bigl( D^{*}_{k_3 k_4} \sigma^{+} a_{k_2,\downarrow} a_{k_1,\uparrow} 
a^{\dagger}_{k_3,\uparrow} a^{\dagger}_{k_4,\downarrow} \sigma^{-} \,\hat{\varrho}_{tot}(t) + 
D_{k_3 k_4} a^{\dagger}_{k_1,\uparrow} a^{\dagger}_{k_2,\downarrow} 
\sigma^{-} \sigma^{+} \, a_{k_4,\downarrow} a_{k_3,\uparrow} \,\hat{\varrho}_{tot}(t) \nonumber\\
&&\hspace{55pt} -D^{*}_{k_3 k_4} \sigma^{+} a_{k_2,\downarrow} a_{k_1,\uparrow} \,\hat{\varrho}_{tot}(t)\, 
a^{\dagger}_{k_3,\uparrow} a^{\dagger}_{k_4,\downarrow} \sigma^{-}   
-D_{k_3 k_4} a^{\dagger}_{k_1,\uparrow} a^{\dagger}_{k_2,\downarrow} \sigma^{-}\,\hat{\varrho}_{tot}(t)\, 
\sigma^{+} \, a_{k_4,\downarrow} a_{k_3,\uparrow}  \nonumber\\
&&\hspace{55pt} -D_{k_3 k_4} \sigma^{+} a_{k_4,\downarrow} a_{k_3,\uparrow} \,\hat{\varrho}_{tot}(t)\, 
a^{\dagger}_{k_1,\uparrow} a^{\dagger}_{k_2,\downarrow} \sigma^{-}  
-D^{*}_{k_3 k_4} a^{\dagger}_{k_3,\uparrow} a^{\dagger}_{k_4,\downarrow} \sigma^{-}\,\hat{\varrho}_{tot}(t)\, 
\sigma^{+} \, a_{k_2,\downarrow} a_{k_1,\uparrow}  \nonumber\\
&&\hspace{55pt} \hat{\varrho}_{tot}(t) \,D_{k_3 k_4} \sigma^{+} a_{k_4,\downarrow} a_{k_3,\uparrow} 
a^{\dagger}_{k_1,\uparrow} a^{\dagger}_{k_2,\downarrow} \sigma^{-} +
\hat{\varrho}_{tot}(t) \, D^{*}_{k_3 k_4} a^{\dagger}_{k_3,\uparrow} a^{\dagger}_{k_4,\downarrow} 
\sigma^{-} \sigma^{+} \, a_{k_2,\downarrow} a_{k_1,\uparrow} \Bigr), \nonumber
\end{eqnarray}   
where 
\begin{eqnarray}\label{D-ftn}
D_{k_1 k_2}=
\pi\hbar \, \delta (\varepsilon^{n}_{k_1}+\varepsilon^{n}_{k_2} - 2\mu -2 E^{(n)}_{Q}) 
+\frac{i \hbar (\varepsilon^{n}_{k_1}+\varepsilon^{n}_{k_2} - 2\mu -2 E^{(n)}_{Q})}{(\varepsilon^{n}_{k_1}+\varepsilon^{n}_{k_2} - 2\mu -2 E^{(n)}_{Q})^2+\delta^2}. 
\end{eqnarray}  
  
We carry out the trace over the states of the normal metal in Eq.~\eqref{Liouville2} 
and find the equation for $\hat{\varrho} = Tr_n [\hat{\varrho}_{tot}]$ as 
\begin{eqnarray}
\partial_t \hat{\varrho}(t) &=& -\frac{i}{\hbar}\left[ H_{CPB}, \hat{\varrho} (t) \right] 
- \check{\hat{\mathfrak{L}}}(\hat{\varrho}),
\end{eqnarray}  
where $\check{\hat{\mathfrak{L}}}(\hat{\varrho})$ describes the effect of the normal metal on the CPB obtained by
integrating out the normal states from Eq.~\eqref{Liouville3} and using the real part of $D_{k_1 k_2}$ in Eq.~\eqref{D-ftn} 
\begin{eqnarray}\label{Lindblad Eq}
\check{\hat{\mathfrak{L}}}(\hat{\varrho})= \gamma(t) \left(\hat{\varrho}(t)+\frac{1}{2}\hat{\sigma}_3 \{\hat{\sigma}_3,\hat{\varrho}(t)\} - \hat{\varrho}_{eq}\right).
\end{eqnarray}
Here the rate $\gamma(t)$ and the equilibrium density matrix $\hat{\varrho}_{eq}$ are 
\begin{eqnarray}
\gamma(t) &=& e^{-\frac{4A}{\lambda}(1+\text{sin}\omega t)} \frac{2\pi t^2_A \nu^2 E^{(n)}_{Q}}{\hbar} \coth \frac{E^{(n)}_{Q}}{k_B T},\\ 
\hat{\varrho}_{eq}&=&\frac{2}{1+e^{2 E^{(n)}_{Q}/k_{\text{B}}T}} 
\begin{pmatrix}
1 & 0 \\
0 & e^{2 E^{(n)}_{Q}/k_{\text{B}}T}
\end{pmatrix} -\hat{I}
= -\hat{\sigma}_3 \,\text{tanh}\frac{E^{(n)}_{Q}}{k_B T}.
\end{eqnarray}
Note that remaining term with the imaginary part  Im[$D_{k_3 k_4}$] gives rise to the imaginary, non-diagonal elements of the density matrix and leads to the small renormalization of the charging energy, 
\begin{eqnarray}
\check{\hat{\mathfrak{L}}}'(\hat{\varrho}) &=& -\frac{i t^2_A \nu^2}{\hbar^2} 
\int d\varepsilon_1 \int d\varepsilon_2 \, \text{Im}D_{\varepsilon_1 \varepsilon_2}
\left[f(\varepsilon_1) f(\varepsilon_2) + (1-f(\varepsilon_1))(1-f(\varepsilon_2))\right] 
\times 
\begin{pmatrix}
0 & -\rho_{10}\\
\rho_{01} & 0
\end{pmatrix}  \\
&\sim& \mathcal{O}(t^2_A \nu^2). \nonumber
\end{eqnarray}

\section*{B. Average current in the stationary regime}
To analyze the stationary regime and average current at arbitrary
$\Gamma$ it is convenient to use Bloch sphere representation
$\hat\varrho(t)=m_{i}(t) \hat{\sigma_{i}}$. As follows from Eqs.
(6-7) the evolution of the parameters $m_{i}$ is described by the
following equation:
\begin{eqnarray}\label{A1}
\frac{\partial \vec m}{\partial t}=\hat L(t)\vec m-\gamma(t)(
2 \vec{m}+\hat{L}_{3}^{2}\hat{m}-f_{eq}\vec{\textbf e}_3)\\
\hat L(t)=2E_{J}(x(t))\hat L_1+2E_{Q}(x(t))\hat L_3, \nonumber
\end{eqnarray}
here $\vec m=(m_1, m_2,m_3)^T,\, \vec{\textbf e}_3=(0,0,1)^T$ and
matrices  $\hat L_i\, (i=1,2,3)$ takes a form,
\begin{equation}\label{LL}
\hat L_1=\left(\begin{array}{cccccc}
  0 & 0 & 0\\
  0 & 0 & -1 \\
  0& 1 & 0 \\
\end{array}\right),\,
 \hat L_2=\left(\begin{array}{cccccc}
  0 & 0 & 1\\
  0 & 0 & 0 \\
  -1& 0 & 0 \\
\end{array}\right),\,
\hat L_3=\left(\begin{array}{cccccc}
  0 & -1 & 0\\
  1 & 0 &  0\\
  0 & 0 & 0 \\
\end{array}\right).
\end{equation}
Note that the matrices $L_i$ are the basis matrices of the Lie
algebra $\mathfrak{so}(3), \,\left[\hat L_i,\hat
L_j\right]=\epsilon_{ijk}\hat L_k$.

In a stationary regime the the vector $\vec m(t)$ is periodic
function of time, $\vec m(t+T_0)=\vec m(t)$. We look for the
stationary solution of Eq.(\ref{A1}), in a form,
\begin{equation}\label{98}
\vec m(t)=e^{2\Phi(t)\hat L_3}\vec{a}(t),
\end{equation}
where $\Phi(t)= \hbar^{-1}\int_{0}^{t}dt'E_{Q}(x(t'))$. Vector
$\vec{a}$ satisfies the following equation
\begin{eqnarray}\label{A1}
\frac{\partial \vec a}{\partial t}=\hat
L_{in}(x(t))\vec{a}-\gamma(t)(
2 \vec{a}+\hat{L}_{3}^{2}\vec{a}-f_{eq}\vec{\textbf e}_3)\\
\hat L_{in}(\tau)= 2\hbar^{-1}E_{J}(x(t))\left[ \cos(2\Phi(t)) \hat
L_1-\sin(2\Phi(t))\hat L_2\right] \nonumber
\end{eqnarray}
Next, introducing two vectors $\vec{a}^{(1)}
=\vec{a}(0)$,$\vec{a}^{(2)} =\vec{a}(T_{0}/2)$ and taking into
account taking into account that $\gamma(t)=0$ on the time interval
$(0,T/2)$ and $E_{J}(t)=0$ on the time interval $(T_{0},T_{0}/2)$ we
obtain the following equation,

\begin{eqnarray}
  \vec{a}^{(2)} &=& e^{2 \Phi_1 \hat L_3}\hat{Q}\vec{a}^{(1)} \\
  \vec{a}^{(1)} &=& e^{-\hat{\Gamma}}e^{2\Phi_2
  \hat{L}_3}\vec{a}^{(2)}+(1-e^{-2\bar{\Gamma}})f_{eq} \vec{e}_3. \\
  \hat{Q}(\Phi_s) &=& e^{\Phi_s\hat L_3}\left(
\begin{array}{ccc}
1 & 0 & 0 \\
0 & \rho^2-\tau^2 & 2\tau\rho \\
0 & -2\tau\rho & \rho^2-\tau^2 \\
\end{array}
\right) e^{\Phi_s\hat L_3} \\
  e^{-\hat{\Gamma}} &=& \left(
  \begin{array}{ccc}
e^{-\bar{\Gamma}} & 0 & 0 \\
0 &e^{-\bar{\Gamma}} & 0 \\
0 & 0 & e^{-2 \bar{\Gamma}} \\
\end{array}\right)
\end{eqnarray}

From these relations, considering that $\hat{L}_3 \vec{e}_3=0$ and
$e^{\hat{\Gamma}}\vec{e}_{3}=e^{2 \bar{\Gamma}}\vec{e}_{3}$ we find
that,
\begin{equation}\label{A5}
\vec{a}^{1}= e^{(\Phi_1+\Phi_2) \hat L_3}\frac{(e^{2 \bar{\Gamma}}-1)f_{eq}}{e^{\hat{\Gamma}} - \hat Q(\Phi)}\vec{e}_3,
\end{equation}
where $\Phi=\Phi_s+\Phi_1+\Phi_2 =\Phi_s+ \hbar^{-1}\int_{0}^{T_{0}}dt'E_{Q}(x(t')) + 2\pi V_G/V_0$.

The average charge transferred to superconductor $Q$ is given by the
scalar product
$$Q=
\left(\vec{e}_{3},(\hat{Q}(\Phi_s)-1)\vec{a}^{(1)}\right)=\left(\vec{e}_{3}, (\hat{Q}(\Phi)-1) \frac{(e^{2 \bar{\Gamma}}-1)f_{eq}}{e^{\hat{\Gamma}} - \hat Q(\Phi)}\vec{e}_3\right)$$ 
After
straightforward but comprehensive algebraic calculations weobtain
the following quite compact general expression for $Q$
\begin{equation}\label{F}
Q=f_{eq}\frac{\tau^2 \sinh\bar{\Gamma}}{\cosh\bar{\Gamma}-\rho^{2}\cos2\Phi}
\end{equation}

\end{document}